\documentclass[a4paper, 10pt, conference]{IEEEtran} % Use this line for a4
\IEEEoverridecommandlockouts

\usepackage{graphicx}
\usepackage[utf8]{inputenc} 
\usepackage{amsmath}
\usepackage[T1]{fontenc}
\usepackage{caption}
\usepackage{subcaption}
\usepackage[margin=1in]{geometry}
\usepackage{censor,caption}
\title{\LARGE \bf Early Detection for Multiversion Concurrency Control Conflicts in Hyperledger Fabric}

\author{
    \IEEEauthorblockN{Helmi Trabelsi\IEEEauthorrefmark{1}, Kaiwen Zhang\IEEEauthorrefmark{1}}

    \IEEEauthorblockA{\IEEEauthorrefmark{1}ÉTS Montreal\\
        {Department of Software and IT Engineering} 
    \\\{helmi.trabelsi.1@ens., kaiwen.zhang@\}etsmtl.ca}

}

\begin{document}
\maketitle

%%%%%%%%%%%%%%%%%%%%%%%%%%%%%%%%%%%%%%%%%%%%%%%%%%%%%%%%%%%%%%%%%%%%%%%%%%%%%%%%
\begin{abstract}
Hyperledger Fabric is a popular permissioned blockchain system that features a highly modular and extensible system for deploying permissioned blockchains which are expected to have a major effect on a wide range of sectors. Unlike traditional blockchain systems such as Bitcoin and Ethereum, Hyperledger Fabric uses the EOV model for transaction processing: the submitted transactions are executed by the endorsing peer, ordered and batched by the ordering services, and validated by the validating peers.
Due to this EOV workflow, a well-documented issue that arises is the multi-version concurrency control conflict. This happens when two transactions try to writes and read the same key in the ledger at the same time. Existing solutions to address this problem includes eliminating blocks  in  favor  of streaming transactions, repairing conflicts during the ordering phase, and automatically  merging  the  conflicting transactions using CRDT (Conflict Free Replicated Data) techniques. 

In this paper, we propose a novel solution called \textit{Early Detection for MVCC Conflicts}. Our solution detects the conflicting transactions at an early stage of the transaction execution instead of processing them until the validation phase to be aborted. The advantage of our solution is that it detects conflict as soon as possible to minimize the overhead of conflicting transaction on the network resulting in the reduction of the end-to-end transaction latency and the increase of the system's effective throughput.
We have successfully implemented our solution in Hyperledger Fabric. We propose three different implementations which realize early detection. Our results show that our solutions all perform better than the baseline Fabric, with our best solution SyncMap which improves the goodput by up to 23\% and reduces the latency by up to 80\%. %at a conflict rate of 40\%.

\end{abstract}

\begin{IEEEkeywords}
Hyperledger Fabric, MVCC, EOV.
\end{IEEEkeywords}

%%%%%%%%%%%%%%%%%%%%%%%%%%%%%%%%%%%%%%%%%%%%%%%%%%%%%%%%%%%%%%%%%%%%%%%%%%%%%%%%
\section{INTRODUCTION}

Hyperledger Fabric is a popular permissionless system that allows the development of blockchain applications in a variety of domains such as finance, healthcare, and supply chain management. Compared to Ethereum \cite{Buterin2014} and Bitcoin \cite{Kornmesser2008}, Fabric offers superior performance, due to the consensus algorithm used and the small number of peers on the network compared to public blockchains. Furthermore, Fabric provides a flexible framework for managing responsibilities across parties using the MSP (membership service provider) which is an abstract component that maintains the identities and roles of all nodes that belong to the same organization.  

Most existing blockchain platforms such as Ethereum and Hyperledger Fabric allow Turing-complete computations by executing a smart contract for a given transaction \cite{Zheng2020AnPlatforms}. 
To guarantee consistency across the network, peers execute transactions to generate the next state of the blockchain after the content of the next block is known (which transactions to execute, and in what order): this is called the Order-Execute (OX) transaction flow \cite{Androulaki2018}. The weaknesses of the OX pattern is that the sequential execution of the transactions within each block limits the throughput of the system. Furthermore, each peer requires knowledge of all smart contracts to compute the next state of the blockchain, which may present confidentiality and privacy issues. 

Instead, Fabric employs the Execute-Order-Validate (EOV) pattern, where the transactions are executed in a sandbox called endorsing peers to generate the read-write sets including the versions of the used keys for the transaction's simulation, then the transactions are ordered by the ordering services into blocks, and at the end validated and committed to the ledger. EOV overcomes the limitations of OX by providing parallelism of transactions execution on different endorsing peers.

The drawback of the EOV pattern is that a read-write lock is used to synchronize the execution and validation phases. To solve this issue, Hyperledger Fabric implements a Multi-Version Concurrency Control mechanism to guarantee the consistency of the blockchain. When validating transactions by the validating peers, the versions of the generated read-write sets are compared to the keys’ versions of the ledger to avoid that two transactions attempt to modify/read the same key/value pairs at the same time. A transaction will be rejected if its read set contains an old version of a key and the client has to resubmit it.  

The aforementioned MVCC (Multiversion Concurrency Control) conflict problem is a well-documented limitation of Hyperledger Fabric as it decreases the effective throughput of the system (also known as goodput). This is because the blocks could contain aborted transactions which still count against the size limit, thereby wasting valuable block space. These aborted transactions must also be retried by their respective clients as brand new transactions, which generates additional load on the system. In practice, the MVCC problem can have a serious impact on the performance of the blockchain, as recent studies show that in realistic scenarios such as electronic health records, 40\% of the transactions failed due to concurrency conflict \cite{Chacko2021WhyVersion}. Prior works seek to address or mitigate the MVCC problem by using a lockless approach to provide transaction isolation \cite{Meir2019} or reordering transactions at the ordering phase to minimize the conflict rate when validating and committing transactions to the ledger \cite{Sharma2019} \cite{Ruan2020}. 

In our paper, we present a novel approach to solve the MVCC problem called \textit{Early Detection of MVCC Conflicts} (EMVCC). We introduce an EMVCC detection mechanism that aims to reduce the number of conflicts between transactions which increases the overall system goodput. The advantage of our solution over existing works is that the MVCC conflict is detected at the first contact of the transaction with the blockchain network at the endorsement policy allowing to improve the network performances.

The contributions of our paper are as follows:
\begin{enumerate}
    \item We provide the formulation of the problem of early detection of conflicting transactions at the endorsement phase, rather the traditional approach of handling MVCC conflicts at the validation phase (Section \ref{sec:pbform}).
    
    \item We propose a novel solution called Early Detection of MVCC Conflicts using local caching at endorsing nodes (Section \ref{sec:propsol}). We provide a theoretical analysis of our solution to calculate the expected effectiveness of the solution depending on parameters such as the endorsement policy, number of organisations, etc. (Section \ref{sec:thAnalysis}).
    
    \item We present three reference implementations of our solution: SyncMap, Lock-Free, and Mutex Lock, their main difference is the data structure used for information storing (Section \ref{sec:datastruc}).
    
    \item We evaluate our solution and compare it to the baseline and we did a sensitivity analysis to study the impact of different network and load parameters on the performances (Section \ref{EXPERIMENTS}).
\end{enumerate}

The paper now continues with Section \ref{BACKGROUND} that gives an overview of the Hyperledger Fabric. This is followed by Section \ref{RELATEDWORK} where we describe prior works related to the MVCC problem.

\section{BACKGROUND}\label{BACKGROUND}

In this section, we give an overview of Hyperledger Fabric Components, we detail the transaction flow and we present the endorsement policies and Multi-Version Concurrency Control Read-Write Conflict (MVCC).  

\subsection{Hyperledger Fabric Components}
Hyperledger Fabric is a permissioned distributed ledger system specialized for business applications. Its modular architecture makes blockchain solutions confidential, resilient, and flexible. Hyperledger Fabric has multiple fundamental components such as the client and the peers which are the endorsing peers and the committing peers. In this paper, we will focus on these components as they are the only components affected by our solutions.  

\subsubsection{Clients}
Clients are applications or software that operate on behalf of a user to submit transactions on the Fabric network through the Hyperledger Fabric SDK.

\subsubsection{Peers}
Peers are a fundamental component of the network because they keep records of the network’s state and a copy of the ledger. Each peer on the network belongs to an organization that may have one or more peers. The administrator can create, start, stop or reconfigure the peer. To be part of the network, the node should join the shared channel between all the participants. There are two types of peers, endorsing peers and validating/committing peers.  In a typical scenario, every Fabric peer typically fulfill both roles of endorsing and validating.
%so the endorsing peer is a special validating peer.

\textbf{Endorsing peers}: this type of node hosts the smart contract executed to simulate transactions and endorse them using the peer signature. When installing the chaincode (smart contract), the administrator specifies which peers will endorse the transactions by defining the endorsement policy rules.

\textbf{Validating/Committing peers}: they are responsible for the VSCC (Validation System Chaincode) used to validate the endorsement policies and the MultiVersion Concurrency Control (MVCC) validation to ensure that the version of the keys read during the endorsement phase is still the same in the ledger to guarantee the deterministic execution of the transaction. Then the committing peers add blocks to the shared ledger and update the blockchain world state. 

\subsection{Transaction Flow}
\begin{figure*}[ht]
\centering
\includegraphics[width=\textwidth]{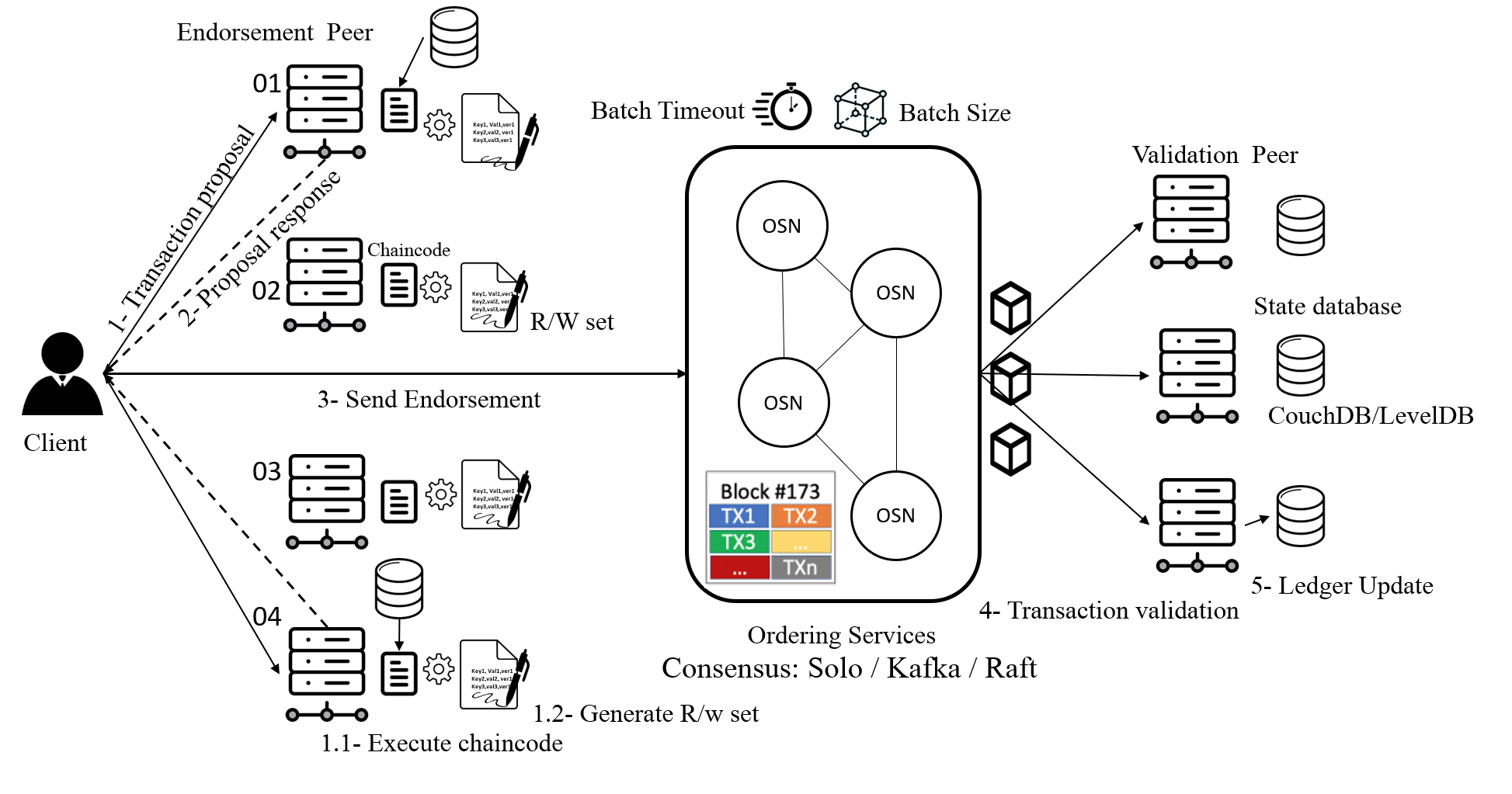}
\caption{Transaction Flow}\label{fig:Figure1}
\end{figure*}

Fig.\ref{fig:Figure1} shows the basic transaction flow in Hyperledger Fabric. It consists of five steps:

\begin{itemize}

    \item \textbf{Step 1}: A client who wants to make a transaction sends a transaction proposal containing the chaincode function and arguments that the client wants to invoke, to the endorsing peer according to the chaincode endorsement policies (see the next Section \ref{sec:epolicy}).
    
    \item \textbf{Step 2}: An endorser receives the transaction proposal, executes the chaincode, and generates the read/write sets which are the sets containing the values and versions of the keys which are read or written while executing the transaction, then the peer creates the proposal response, signs it and sends the response to the client.
    
    \item \textbf{Step 3}:  Once the client collects the required number of endorsements as defined in the endorsement policies, he sends the transaction to the ordering services, this operation contains details about the initial proposal as well as all peer endorsements and read/write sets. The ordering services order transactions received from clients into a block considering the block BatchSize and block BatchTimeout, then the block is created and delivered to the validating peer.
    
    \item \textbf{Step 4}: Upon receiving the block, the committing peer iterates over all the transactions within the block to perform the syntax validation of each transaction, the VSCC validation, and the MVCC validation (see the next Section \ref{sec:mvcc}).
    
    \item \textbf{Step 5}: If a transaction is marked as valid after passing the three checks, it will be added to the new block and the ledger is updated by applying the transaction write sets, else invalid transactions will be rejected.
    
\end{itemize}

\subsection{Endorsements Policies} \label{sec:epolicy}
Hyperledger Fabric allows developers to set policies at the chaincode level. The endorsement policies are rules which specify which peers can agree on the transaction execution before it is added to the ledger. In general, endorsement policies are configured while installing the chaincode and can be modified only during a chaincode upgrade. Once the client creates the transaction, it sends a transaction proposal to all endorsing peers that satisfy the endorsement policy \cite{Manevich2018ServiceFabric} and waits for the proposal responses. When the client receives enough responses and signatures to satisfy the endorsement policy of the chaincode, it can submit the transaction with the endorsement signatures to the ordering services. Endorsement policies can be defined as follow: 

\begin{itemize}
    \item \texttt{AND('Org1.member','Org2.member')}: the client needs one signature from each organization to be able to submit a transaction
    \item \texttt {OR('Org1.member','Org2.member')}: the client needs one signature from either one of the two organization to be able to submit a transaction
    \item \texttt{OR('Org1.member',AND('Org2.member' ,'Org3.member'))}: the client needs either one signature from a member of the Organization1 or one signature from a member of the Organization2 and one signature from a member of the Organization3
    \item \texttt {Out-Of(2,'Org1.member', 'Org2.member','Org3.member')}: At least two of the three organization must endorse the transaction to be valid. It is equivalent to \texttt{OR(AND('Org1.member','Org2.member') ,AND('Org1.member','Org3.member'), AND('Org2.member','Org3.member'))}
\end{itemize}

In the scope of this paper, we study the first two examples (AND, OR) for our solution, however types 3 and 4 have been studied but are not included in this paper due to lack of space. Furthermore, we assume that each client will select endorsing peers randomly if given the choice in accordance to the policy.

\begin{figure*}[htbp]
\centering
\includegraphics[width=0.6\textwidth]{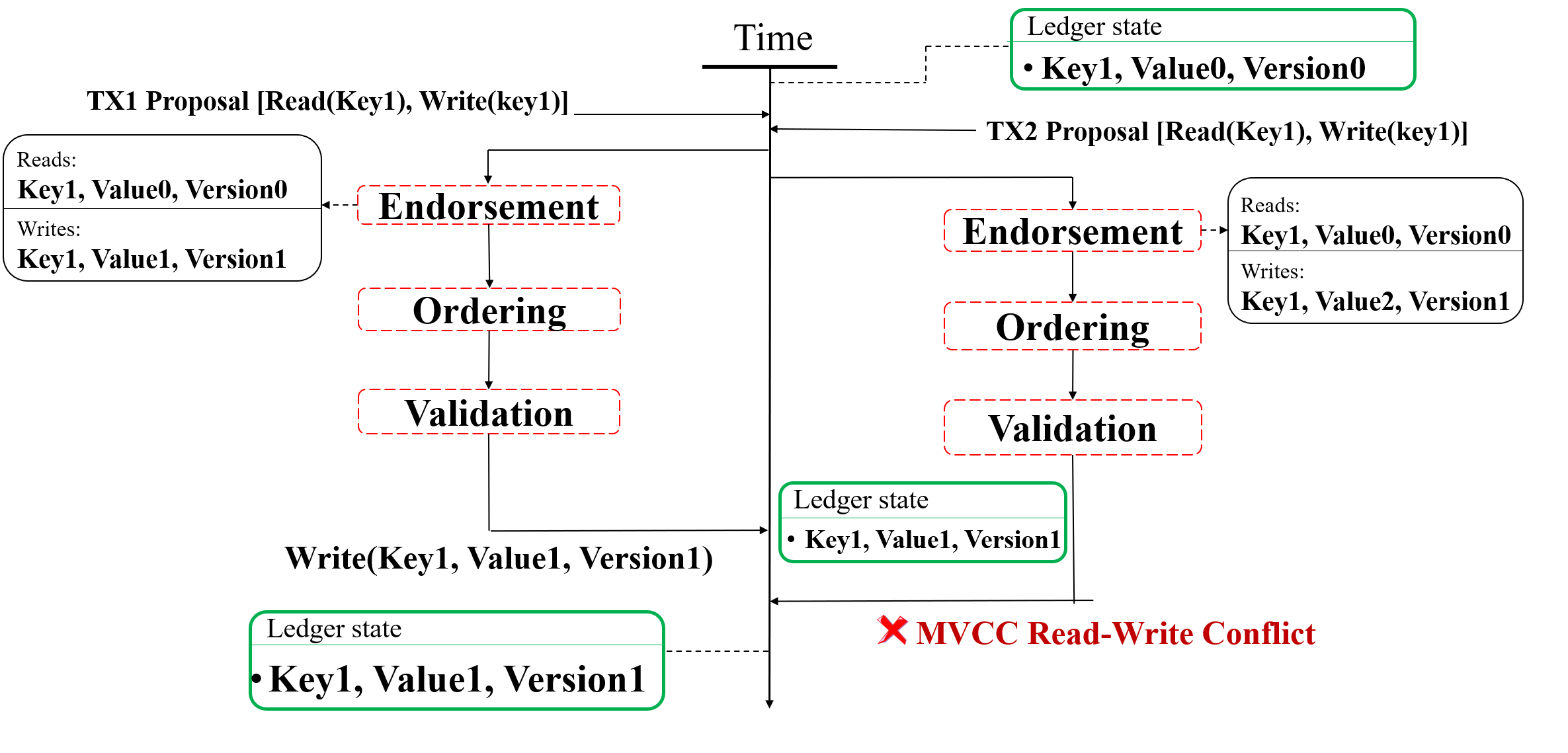} 
\caption{MVCC Read-Write Conflict Example} \label{fig:Figure2}
\label{fig}
\end{figure*}

\subsection{Multi-Version Concurrency Control Read-Write Conflict (MVCC)}  \label{sec:mvcc}

Hyperledger Fabric uses a Multi-version concurrency control system to ensure the consistency of the ledger. This mechanism validates that the versions of keys read at the endorsement time of the transaction are still the same at the validation phase \cite{Larson2012High-PerformanceDatabases}. This process guarantees that there are no reads of old values that have been changed by another concurrent transaction. During the time period between the endorsement and the validation phase, if another transaction has updated the version of the keys listed in the read set, the transaction will fail at the MVCC validation.
 
The multi-version concurrency control read-write conflict is a problem that occurs when two clients try to update and read the same key at the same time. Fig.\ref{fig:Figure2} shows an example of this conflict. Let’s assume that  \textit{user1} submits  \textit{TX1}, and at the same time \textit{user2} submits \textit{TX2}, the two transactions will read and update the same value \textit{Value1} of \textit{Key1}. After endorsing and ordering simultaneously the transactions, \textit{TX1} will be validated and its write set will be applied to the ledger resulting in the modification of the value and version associated with \textit{Key1} to \textit{value1}  and \textit{version1}. However, when \textit{TX2} underwent the MVCC validation, it fails because the version of \textit{Key1} in the read set of the transaction is not the same in the ledger. Therefore, the MVCC validation detects the inconsistency between the ledger and the endorsement result and return an MVCC error.

\section{RELATED WORKS}\label{RELATEDWORK}

Hyperledger Fabric is a relatively new system that is already experiencing some major architectural improvements. The majority of related work aims to improve the throughput and minimize the latency of the network, but there is a lack of effective solutions to deal with the MVCC problem and in most cases, the transaction conflict factor is not considered in the evaluation results. In this section, we review recent research on techniques to improve the Hyperledger Fabric performances. We will review this works along three categories: works that optimize endorsement phase, works that improve the ordering phase, and works that enhance the validation phase.

\subsection{Endorsement Phase Optimization}
%We can classify works that optimize the endorsement phase into three groups. The first set represents the solutions that affect the endorsement policies.

The work done by \cite{Kwon2019} is aiming to improve the read transaction processing by distinguishing between reading and writing transactions during the endorsement process. As a result, the transaction endorsement latency is reduced by 60\% compared to the traditional fabric network. This approach is complementary to our own solution since early detection can still be applied to further reduce the latency of failed transactions.

\subsection{Ordering Phase Optimization}
For the ordering phase, \cite{Istvan2018} propose the elimination of the concept of blocks in favor of processing transactions in streaming to decrease the batching overhead. Also, the authors implement an FPGA-based (Field Programmable Gate Array) consensus for the ordering service that decreases the commit latency below a millisecond by cutting latency in half compared to the Raft-based ordering service. 

The authors of Fast Fabric \cite{Gorenflo2019} redesign the fabric ordering service to work with only the transaction IDs by Separating the transaction header from the payload to decide the transaction order only with the transaction’s IDs which makes transaction processing in the ordering services faster which increase the throughput. 

\cite{Sharma2019} propose Fabric++ a solution that aims to reduce the MVCC failure rate by reordering transactions at the ordering phase using a conflict graph to abort the transaction that cannot be serialized. 

\cite{Ruan2020} designed an optimized extension of Fabric++ that can handle both inter-block and intra-block MVCC read conflicts.

Our solutions detect the MVCC conflict at the endorsement phase, however, Fabric++ and FabricSharp aim to reorder transactions in the ordering phase to avoid the maximum number of conflicts.

\subsection{Validation Phase Optimization}
Multiple articles  propose the parallel execution of the validation process (syntax verification, endorsement policy verification, MVCC validation) to accelerate the block validation \cite{Gorenflo2019a,Gorenflo2019,Istvan2018,Javaid2019,Thakkar2018}. 

FabricCRDT focuses on automatically merging the conflicting transactions using CRDT techniques without rejecting them~\cite{Nasirifard}. 
 \cite{Istvan2018} introduce a disk writes batching mechanism by using a local batcher that accumulates the ledger update operations until reaching a batch size or batch time out to write to the database.

%As an extension to batching writes, \citeauthor*{Thakkar2018} \cite{Thakkar2018} propose bulk read operations during the MVCC validation to make a single API call per block to the database. 

Finally, \cite{Meir2019} presented a lock-free solution for providing transaction isolation, this approach allows the removal of the shared lock while ensuring transaction isolation. 

The proposed improvements of the validation phase are based on the parallelization of validation processes or the optimization of reading and writing operations on the databases. All of these works offer useful insights into various techniques that can improve the Hyperledger Fabric performance and represent a complementary improvement to our proposed solution.

\subsection{MVCC in database systems}
MVCC is one of the most studied problems in the database community which is reused in blockchain systems. This problem is addressed by several works such as the work of \cite{Larson2012High-PerformanceDatabases} who propose two optimized MVCC mechanisms for in-memory databases. Compared to single-version locking, the proposed schemes have higher overhead but are much less sensitive to hotspots and the presence of long-running transactions. Also, \cite{Faleiro2015} introduce the BOHM, a concurrency control mechanism for main-memory multi-versioned database systems ensuring that reads operations never block writes. These proposed solutions can serve to improve the databases-based blockchain systems. However, our proposed solution is preventive because it deals with the MVCC problem before it occurs at the validation phase.

\section{EMVCC Detection \& Solution}\label{OURSOLUTION}

\begin{figure*}[t]
\centering
\includegraphics[width=1\textwidth]{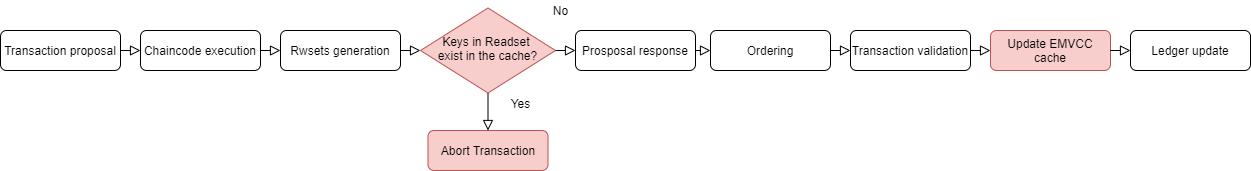} 
\caption{EMVCC Detection Mechanism} \label{fig:Figure3}
\label{fig}
\end{figure*}

In this section, we formulate the problem of MVCC detection at the endorsement phase. We then describe our proposed solutions called early MVCC detection, and present the different possible caching data structures. Also, we present a theoretical analysis of different endorsement policies.     

\subsection{System Model}
In this section, we will use the following model: given an Hyperledger Fabric\footnote{Our solution has been tested with Fabric version 2.0 and higher.} network containing $N$ organizations where each organization has $M$ peers and $M_i$ is the number of peers for organisation $i$. The system uses $Nb_{Tx/Block}$ as the number of transactions per block. The ordering services of the system can use any consensus protocol available in Hyperledger Fabric (e.g., Raft, Kafka, Solo). The chaincode executed by the application generates $\%cfct$ as the conflict rate representing the percentage of transactions which abort due to an MVCC conflict error. Intuitively speaking, a chaincode where transactions read and write to a few shared keys will have a higher conflict rate than that of a chaincode where transactions are independently working on disjoint keys. In order to store chain state, the system can be implemented using any of the available databases compatible with Hyperledger Fabric (CouchDB, LevelDB, etc.).

\subsection{Problem Formulation} \label{sec:pbform}
 The problem we are solving is detecting MVCC conflict by filtering the transactions at the \textit{endorsement phase} based on the endorsed transactions history of each endorsing peer. This is in contrast to the current problem resolved by Fabric, which is that of detecting MVCC conflicts at the \textit{validation phase}, which occurs later in the transaction flow.
 
 We start by defining some basic terms that we will use later:
\begin{itemize}
\item \textbf{Probability of non-detection:} This is a metric to evaluate the performance of our solution. A perfect solution would have a $\mathcal{P}(ND)=0$. It represents the probability that our proposed solution does not detect a conflicting transaction at the endorsement phase, even though it has a conflict with another transaction, such that the MVCC conflict is only detected later at the validation phase.

\item \textbf{False positives:} We call a false positive a transaction which is wrongly declared invalid using our solution, but would have been successfully confirmed using the standard approach. This can occur if our solution is overly aggressive in detecting conflicts and will prematurely abort transactions which have a chance to be successful, which would force the client to retry the entire endorsement process with a new transaction. The false positive rate can be calculated as follows:

\begin{equation}\label{eq1}
\resizebox{0.8\hsize}{!}{$%
FP= \frac{Tx/Block \times \%cfct^2 \times \mathcal{P}(ND) \times (1-\mathcal{P}(ND))} {\%cfct \times \mathcal{P}(ND) - \%cfct + 1}
$%
}%
\end{equation}

where ${Tx/Block}$ is the number of transactions per block, $\mathcal{P}(ND)$ is the probability of non-detection and $\%cfct$ is the conflict rate

\item \textbf{False negatives:} It is the opposite error where a transaction is declared valid at the EMVCC validation stage while it has conflicts with another transaction resulting in its failure later on. Ideally, the number of false negatives should be equal to zero.

\begin{equation}\label{eq2}
    FN= TX\_rate \times \%cfct \times \mathcal{P}(ND)
\end{equation}

\item \textbf{Goodput}\cite{Yu2018CoinExpress:Networks} is the rate of successful transactions that the network can write into the blockchain. In contrast, the throughput of the network is the rate of total transactions which passed through the ordering phase, including transactions which will be aborted due to MVCC conflicts. Ideally the goodput should be equal to the throughput. In a standard Fabric implementation, the goodput is directly impacted by the MVCC conflict rate, since any aborted transaction is still part of a proposed block. Our solution can prevent such conflicts from being reflected inside the blockchain by detecting them prior to the ordering phase and thus improve the goodput.

We can calculate it as follows:

\begin{equation}\label{eq3}
\resizebox{0.7\hsize}{!}{$%
goodput=\frac{Nb\_valid\_TX}{Nb\_TX} \times Throughput
$%
}%
\end{equation}
\end{itemize}

\subsection{Proposed Solution}\label{sec:propsol}
In the current implementation of Fabric, a conflicting transaction passes through the network and eventually fails, thereby consuming unnecessary resources. In our proposed solution, we add a layer to filter transactions at the endorsement phase to abort any detected conflicting transactions. We call our approach Early Multi-Version Concurrency Control (EMVCC). 

As described in Fig.\ref{fig:Figure3}, our EMVCC solution operates after Fabric executes the chaincode and generates the read/write sets, and before an endorsement response is sent back to the client. Our EMVCC solution will compare the read set to a list of pending transactions stored in the peer’s local cache. This cache is populated with transactions currently in progress this peer has previously endorsed which have not yet been confirmed (or aborted).

If a transaction reads a key that is stored in the cache, the transaction will be aborted since an instance of early MVCC conflict was detected. If not, the transaction is stored in the local cache and an endorsement is sent as usual back to the client. Once the transaction is validated or rejected at the validating phase, the cache is updated by removing the keys of the write set of this transaction. 

Note that the usual MVCC phase is still performed during the validation phase, since our EMVCC does not detect conflicts perfectly (see Section~\ref{sec:thAnalysis}). This allows the regular validation phase to catch any MVCC conflicts that are not detected through our EMVCC solution, thereby maintaining correctness of the EOV (Execute-Order-Validate) execution flow.
%This added layer does not affect the Fabric consistency because all transactions will pass the MVCC validation at the validation phase.

\begin{figure}[t]
\centering
\includegraphics[width=0.48\textwidth]{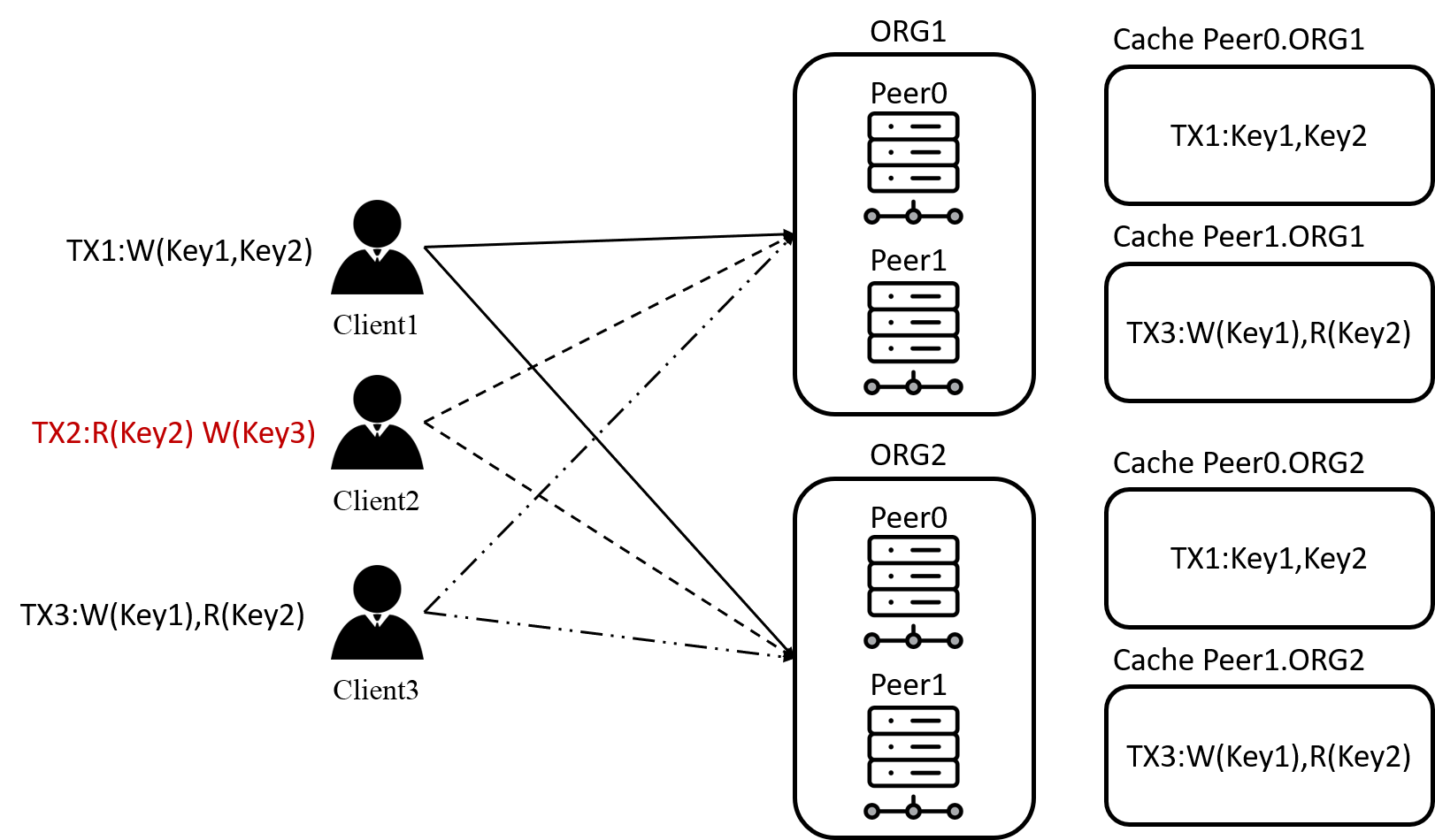} 
\caption{EMVCC Detection Example} \label{fig:EMVCCExample}
\label{fig}
\end{figure}

Fig.\ref{fig:EMVCCExample} shows an example of how EMVCC detection works. Suppose that we have a Fabric network composed of two organizations, each organization having two peers. The chaincode endorsement policy is AND (Org1,Org2). Suppose that \textit{client1} submits \textit{TX1} which writes on \textit{Key1} and \textit{Key2}. This transaction is endorsed by \textit{Peer0.ORG1} and \textit{Peer0.ORG2}. Now suppose we have a transaction \textit{TX2} which reads \textit{Key2} and writes on \textit{Key3}. If this transaction is endorsed by \textit{Peer0.ORG1} or \textit{Peer0.ORG2}, it will be aborted at the endorsement phase because \textit{TX1} is in the EMVCC cache of \textit{Peer0.ORG1} or \textit{Peer0.ORG2}. Suppose another transaction \textit{TX3} arrives which writes on \textit{Key1} and reads \textit{Key2}. If it is endorsed by  \textit{Peer1.ORG1} and \textit{Peer1.ORG2}, it will not be detected by the EMVCC detection mechanism because these peers does not have any conflict with this keys in their EMVCC cache. However, it will still fail later on at the MVCC validation phase.

\subsection{Theoretical Analysis of Non-Detection}\label{sec:thAnalysis}
Since our approach is lightweight in nature and only uses local caching without resorting to any synchronization, it is possible for a conflicting transaction to pass the EMVCC phase undetected. This can happen if the subset of peers chosen to endorse the current transaction have not previously endorsed a pending transaction which conflicts with it. Thus, the endorsement policy, as well as the number of peers and the number of organisations, have a direct impact on the probability of non-detection of our proposed EMVCC solution.

In order to assess the performance of our solution, we present a analytical model which calculates the expected probability of non-detection for our solution using the AND and OR endorsement policies. These theoretical results will be compared to the real results in Section \ref{EXPERIMENTS} to validate the correctness of our implementation. For this analysis, we assume the selection of peers (and organisations) is done uniformly at random when a client is presented with a choice of multiple peers (or organisations) to satisfy the endorsement policy, as it is currently implemented in Hyperledger Fabric \cite{Manevich2018ServiceFabric}. Due to the lack of space, we defer the full details of our mathematical formulas derived here to our extended report.

\subsubsection{Non-Detection in the AND Endorsement Policy}
Using the AND endorsement policy, the probability of non-detection can be calculated as follows:  

\begin{equation}\label{eq4}
\mathcal{P}(ND) = \prod_{n=1}^{N} \frac{(M_i-1)}{M_i}
\end{equation}

where N is the number of organizations and $M_i$ is the number of peers per organization

\begin{figure}[t]
\includegraphics[width=0.45\textwidth]{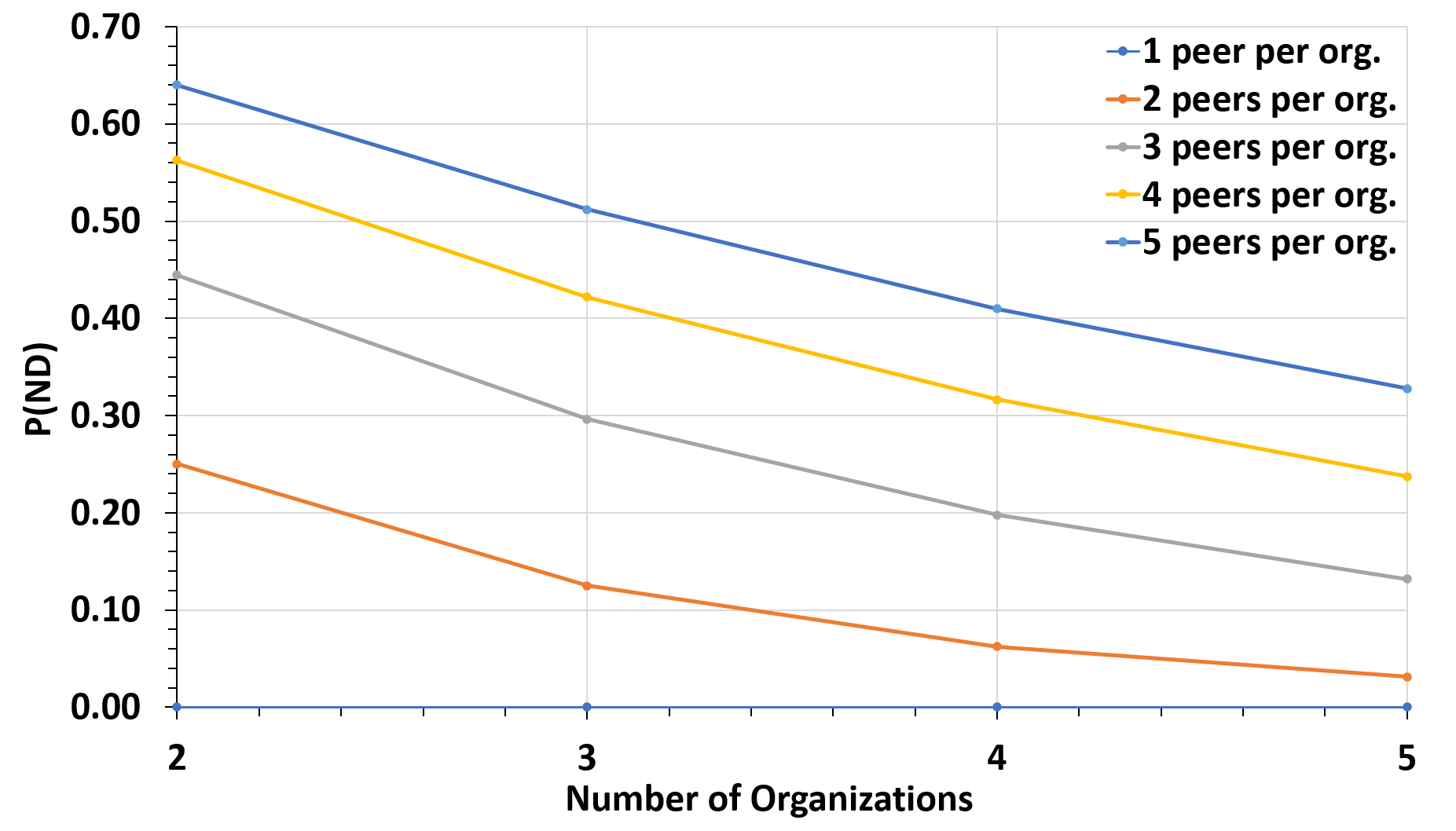} 
\caption{Probability of Non-Detection using AND} \label{fig:FigureAND}
\label{fig}
\end{figure}

Fig.\ref{fig:FigureAND} plots the theoretical evolution of the probability of non-detection of the EMVCC solution for various organizations and peer numbers using AND endorsement policy. We observe that the probability of non-detection decreases when the number of organizations increases, however, it increases when the number of peers per organization increases.

\subsubsection{Non-Detection in the OR Endorsement Policy}
We also analyze the evolution of the probability of non-detection using the OR endorsement policy, it can be calculated using the following formula:

\begin{equation}\label{eq5}
\mathcal{P}(ND)=\left\lbrace
\begin{array}{ll}
\frac{NM-1}{NM}& \mbox{if $M_i$=$M_{i+1}$}\\
\frac{1}{N} {\sum\limits_{i=1}^N \frac{1}{M_i}} & \mbox{else}
\end{array}
\right.
\end{equation}

where N is the number of organizations and $M_i$ is the number of peers per organization

\begin{figure}[t]
\includegraphics[width=0.45\textwidth]{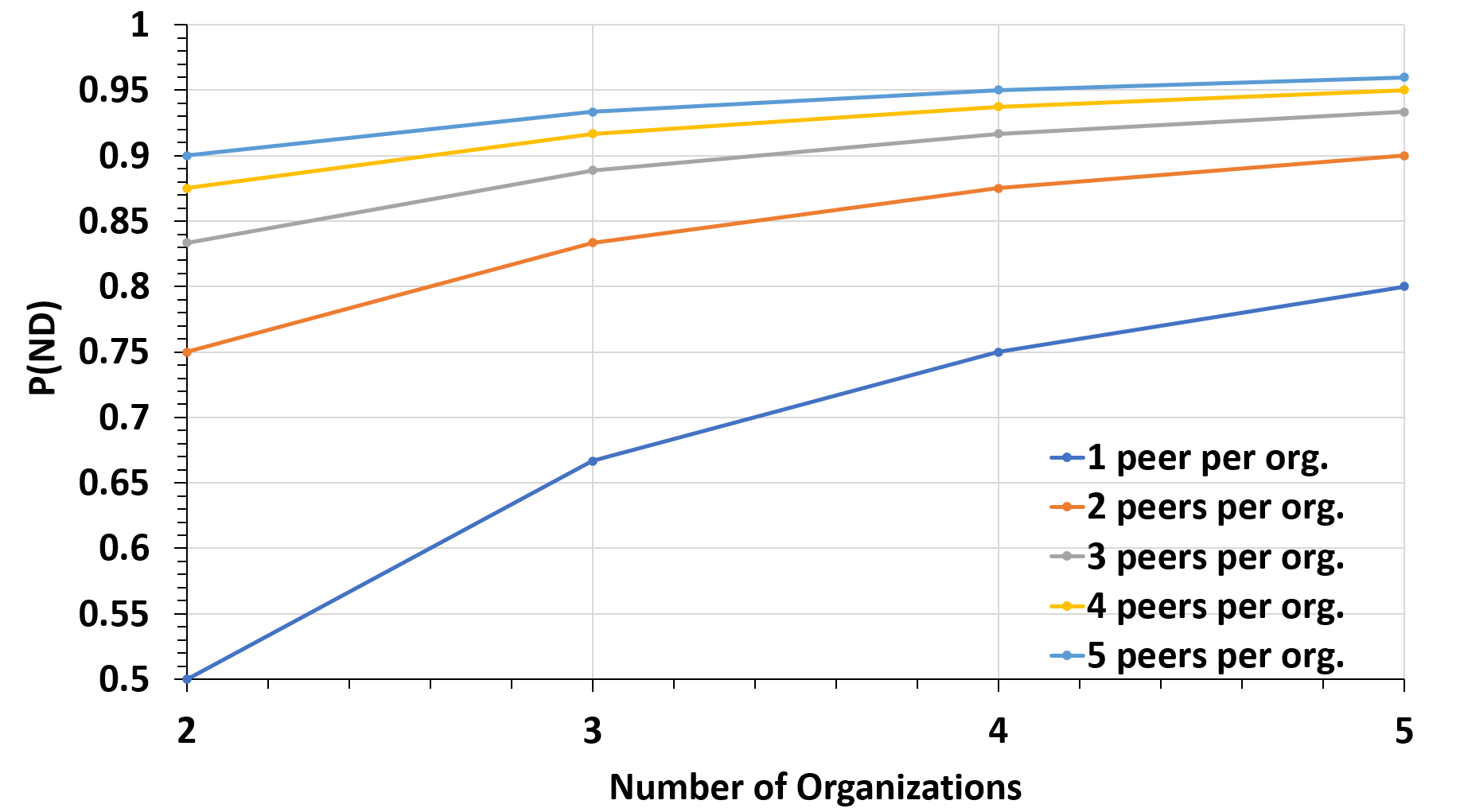} 
\caption{Probability of Non-Detection using OR} \label{fig:FigureOR}
\label{fig}
\end{figure}

Fig.\ref{fig:FigureOR} illustrates the theoretical evolution of the probability of non-detection for different organizations and peers’ numbers using OR endorsement policy. We identify that the probability of non-detection increases when the number of organizations or the number of peers increases due to the lack of synchronization between nodes caches.

\subsection{Choice of Caching Data Structure}  \label{sec:datastruc}
For the local cache maintained by each peer, we propose three different techniques in our reference implementation:

\textbf{MutexLock}: In this implementation, we are using a lock-based data structure to ensure cache consistency while using a multithreaded system \cite{Alexandrescu2007Lock-FreeStructures}. For example, when an endorser reads or writes on the cache, it will be locked and the validator cannot update this entry in the cache. The main disadvantage of this approach is that the locking system adds a delay to the read and write operations since some threads have to wait until a lock is released.

\textbf{LockFree}: With this approach, more than one thread can access the cache concurrently. The read and writes operations are stored in a buffer then once the buffer threshold is reached, the batch is applied to the cache which guarantees that the majority of threads make progress at each step \cite{Alexandrescu2007Lock-FreeStructures}. This technique is useful to decrease the number of times a lock is acquired or released.

\textbf{SyncMap}: Sync Map is an optimized and safe data structure,  which is part of the sync package since Golang 1.9. It uses a dirty map to write new values or updates allowing the read operation to be done without a lock. However, the Mutex locking is essential if numerous concurrent threads (\textit{goroutines}) are writing concurrently in the dirty map and the read-only map is updated by a batch of operations from the dirty map.

\section{BENCHMARKING RESULTS}\label{EXPERIMENTS}
In this section, we evaluate the effectiveness of our proposed EMVCC solution using three different cache implementations: Mutex Lock, LockFree, and SyncMap. We first compare our three different implementations against the standard Fabric implementation as a baseline. We then perform a sensitivity analysis to study the influence of certain system parameters and workload characteristics on the system performance. Table \ref{tab:tab1} shows the default used parameters for our experiment. We use 500 transactions per block as the block size and a block time of 2 seconds which are the ordering parameters that lead to the highest throughput from our experiential testing (see Section~\ref{subsec:sa}). Furthermore, we use a conflict rate of 40\% because in realistic scenarios it represents the percentage of failed transactions due to concurrency conflict \cite{Chacko2021WhyVersion}. We chose the Raft consensus over the the two alternative ordering services because the solo mode is recommended only for development and testing purposes \cite{Wang2020PerformanceFabric} and Kafka is deprecated in versions 2.x of Fabric. We also chose 3 ordering services nodes to satisfy the trade-off between the cost of the network and its performance.

\subsection{Test Environment Characteristic}
Our cluster consists of six E2 virtual machines hosted on the Google Cloud Platform. Two machines serve as two peers and three machines running the ordering services with a raft consensus algorithm. The other virtual machine serves as clients by running the Hyperledger Caliper \cite{Wickboldt2019BenchmarkingSystem} benchmarking tool. Each virtual machine has 8 vCPUs (Virtual Central Processing Unit) and 32 GB of RAM. All the virtual machines are running Ubuntu 16.04 as an operating system. The Fabric peers are set up to use our modified version of Hyperledger 2.3 images and CouchDB as the state database. For the chaincode, we use the Fabcar chaincode which allows us to create a car on the blockchain and modify its owner. 

\subsection{Performances Metrics}
The main metrics for our benchmarking are:
\begin{enumerate}
    \item \textbf{Goodput}: the effective throughput of committed transactions written to the blockchain excluding the aborted ones (Section \ref{sec:pbform}).
    \item \textbf{Latency}: the time between the initial request by the client of the transaction and its final commit to the ledger.
    \item \textbf{EMVVC vs. MVCC rate}: the percentage of rejected transactions due to an EMVCC error versus the percentage of rejected transactions due to an MVCC error.
    \item \textbf{TD EMVCC and TD MVCC}: the time-to-detect is the total time duration between the submission of the transaction by the client and its rejection by an MVCC or EMVCC error.
\end{enumerate}

\begin{figure*}
\begin{subfigure}[t]{0.45\textwidth}
\centering
\includegraphics[width=\linewidth]{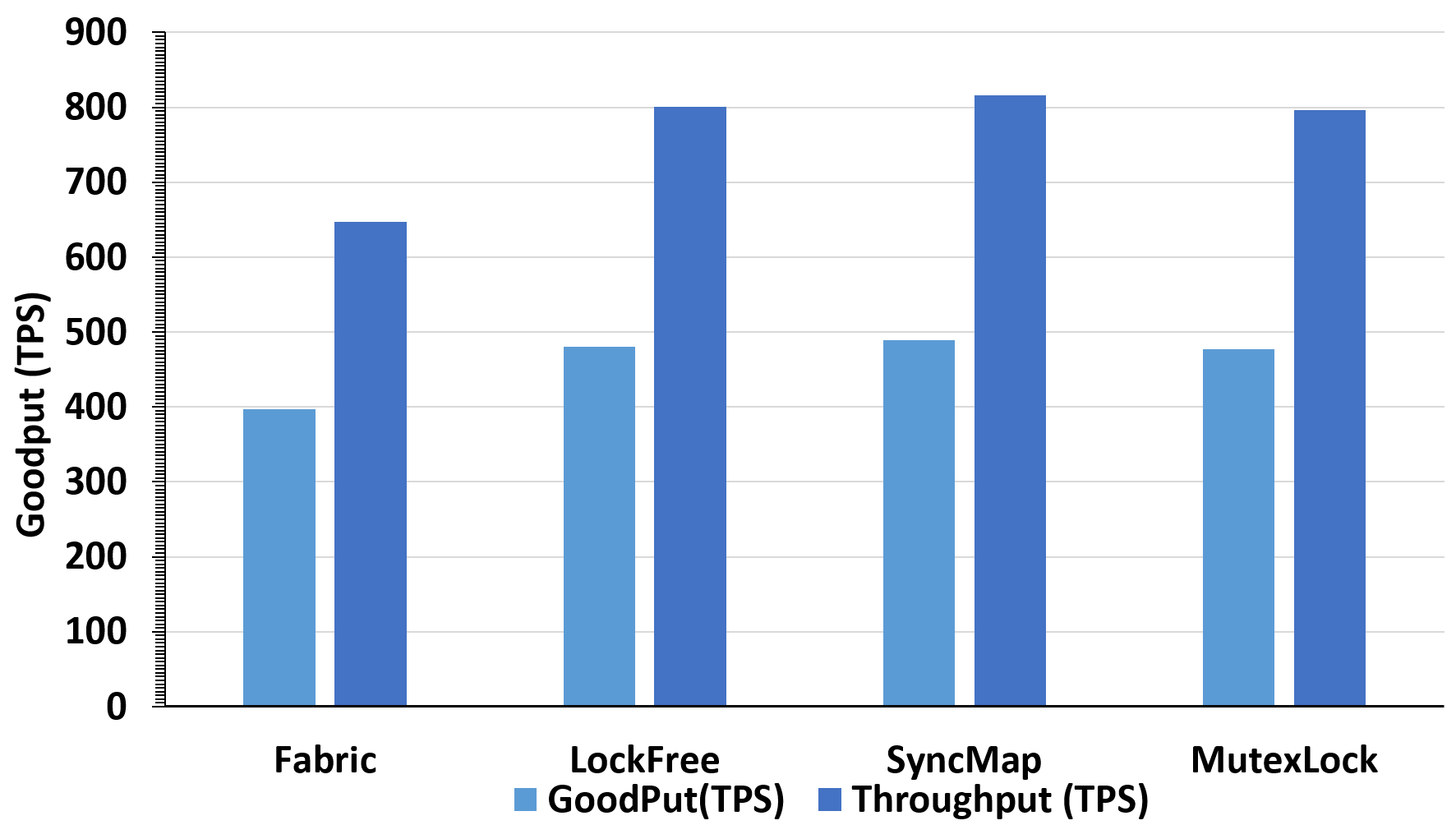}
\caption{Throughput Comparison With The Baseline Solution} 
\label{fig:Comparaison}
\end{subfigure}
\begin{subfigure}[t]{0.45\textwidth}
\centering
\includegraphics[width=\linewidth]{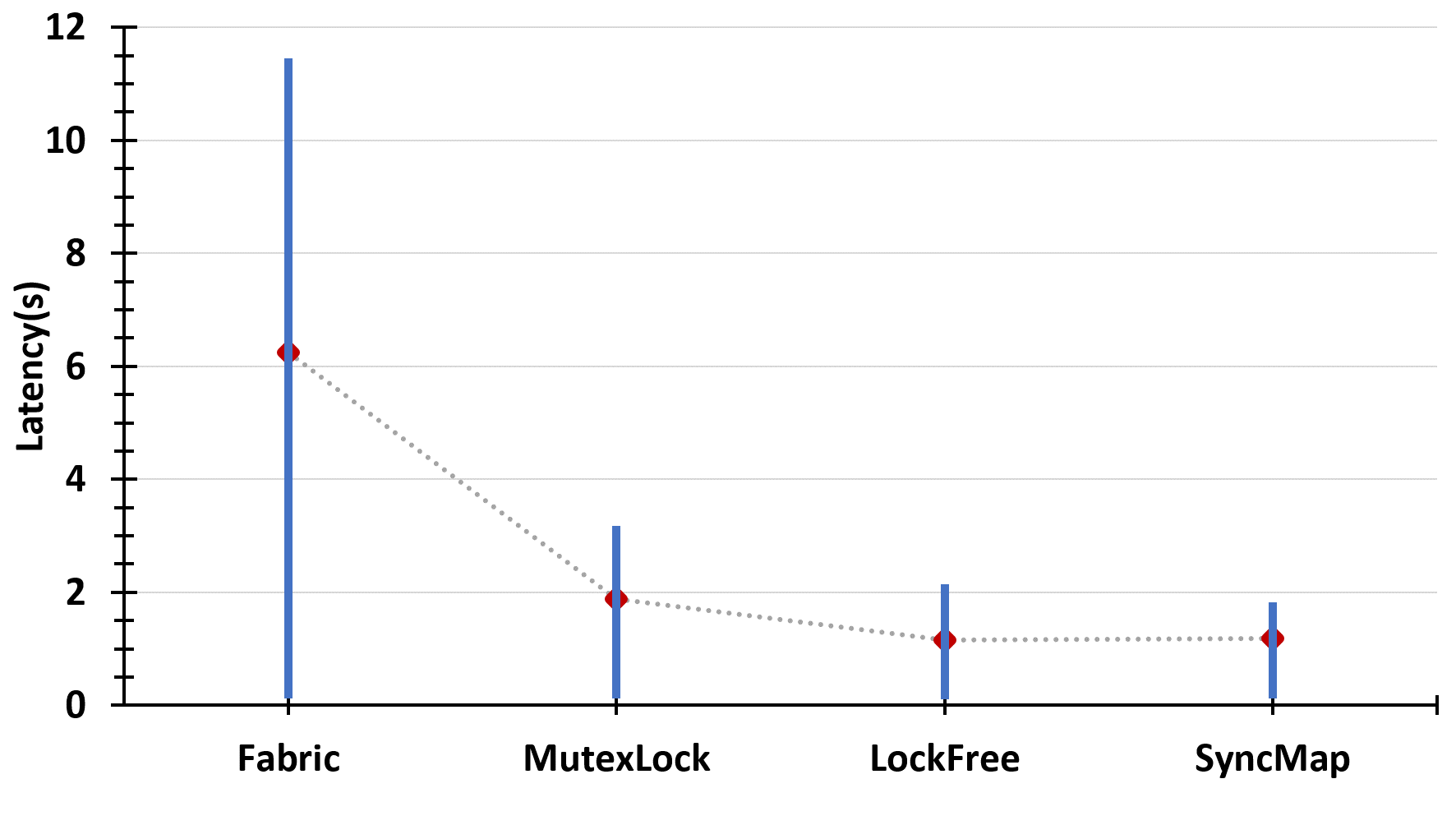}
\caption{Latency Comparison With The Baseline Solution} 
\label{fig:latency}
\end{subfigure}
\caption{Comparison With The Baseline Solution}
\end{figure*}

\subsection{Comparison With The Baseline Solution}
In this section, we compare our proposed  solutions with version 2.3 of Hyperledger Fabric. The throughput and transaction latency are averaged over ten runs.

\begin{table}[]
\centering
\caption{Default Evaluation Parameters} \label{tab:tab1}
\begin{tabular}{|l|l|}
\hline
\textbf{Parameters}         & \textbf{Values} \\ \hline
\textbf{Block size}         & 500 TXs             \\ \hline
\textbf{Block time}         & 2 seconds             \\ \hline
\textbf{Number of organizations}         & 2 \\ \hline
\textbf{Number of peers per organization}         & 2 \\ \hline
\textbf{Ordering Consensus}         & Raft (3 Orderers) \\ \hline
\textbf{MVCC Conflict rate} & 40\%            \\ \hline
\textbf{Endorsement Policy} & AND(Org1,Org2)            \\ \hline
\end{tabular}
\end{table}

Fig.\ref{fig:Comparaison} plots the average throughput for three implementations and the Hyperledger Fabric. The three proposed solutions are better than the Fabric baseline, the best solution being SyncMap with a 23.2\% of goodput improvement compared to Fabric. For Mutex Lock and LockFree, the percentage of improvement is 20.4\% and 21.2\%, respectively. Also, Fig.\ref{fig:latency} plots the average latency for the three implementations and the baseline. Similarly, the three implementations reduce latency compared to the baseline: for SyncMap and LockFree, the latency is reduced by 80\%, and for MutexLock by 69\%. This is due to the early detection of the conflicting transaction at the endorsement instead of going until the validation phase to be aborted. We thus conclude that SyncMap is the best data structure for our early MVCC cache implementation because it is the best for read operations when the the peer has multiple vcPUs.

\subsection{Sensitivity Analysis}
\label{subsec:sa}

In this section, we analyse the impact of various network and workload parameters on the performance metrics such as the conflict rate, endorsement policies, and compute resources per peer, etc. 

\subsubsection{Impact of Conflict Rate}
The conflict rate is the most important parameter that can show us the utility of our proposed solution. In Fig. \ref{fig:conflict}, we plot the average goodput for Fabric and our different solutions over different values of conflict rates. As expected, with an increase in the conflict rate, the goodput decreases because we are increasing the number of failed transactions until we reach a 100\% conflict rate where the goodput becomes zero. Also, we can see that the three proposed solutions perform better than Fabric for different values of the conflict rate. SyncMap is the best one, since it is improving the goodput by 10\% when the conflict is around 20\%  and its impact is more important by exceeding 20\% to reach an improvement of 23\% at a conflict rate of 40\%. The relative performance of this solution increases over the baseline as the conflict rate increases for an application.

\subsubsection{Impact of Compute Resources per Peer}

\begin{figure*}[h!]
\begin{subfigure}[t]{0.30\textwidth}
\centering
\includegraphics[width=\linewidth]{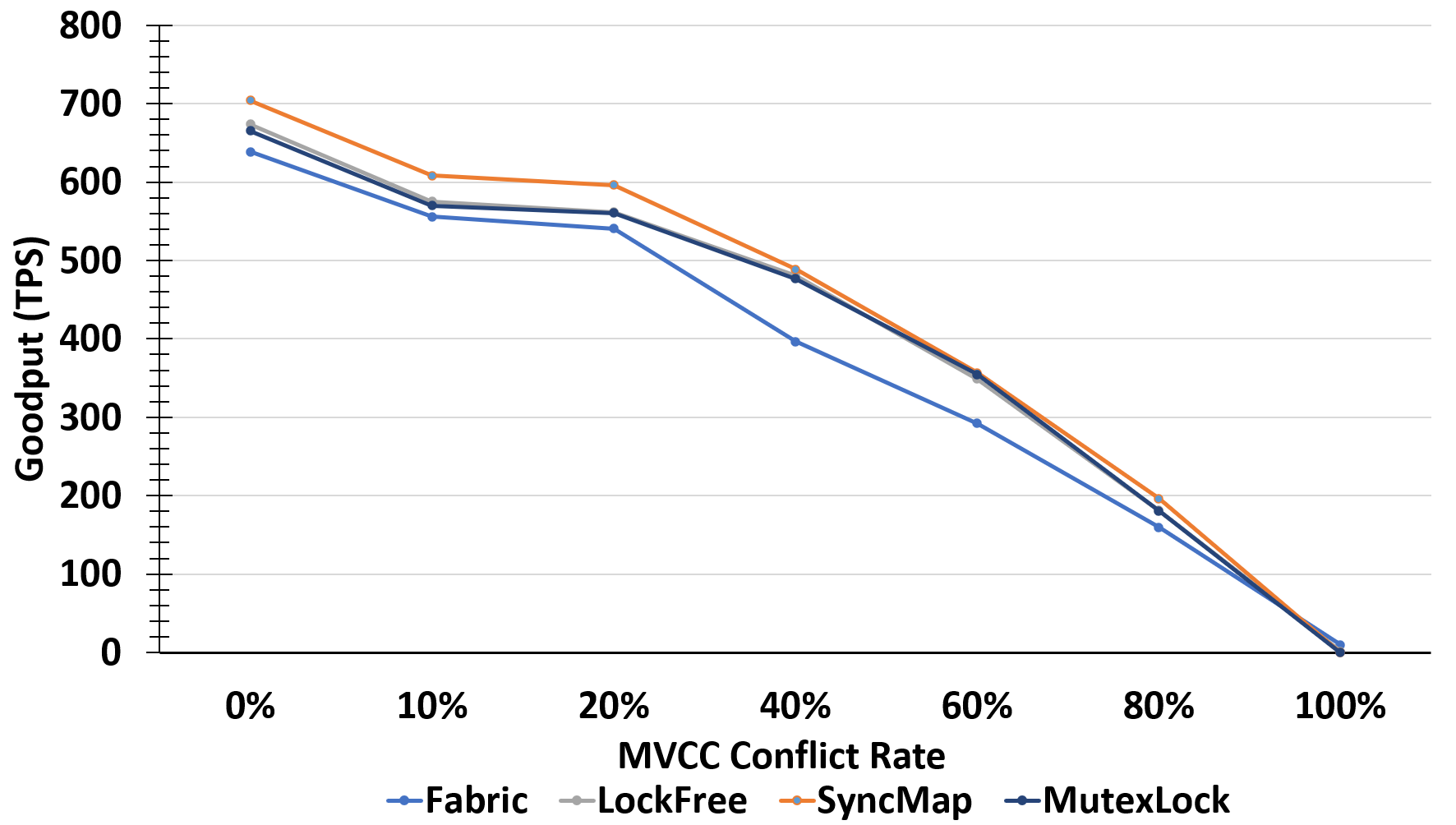}
\caption{MVCC Conflict Rate vs. Goodput} 
\label{fig:conflict}
\end{subfigure}
\centering
\begin{subfigure}[t]{0.30\textwidth}
\centering
\includegraphics[width=\linewidth]{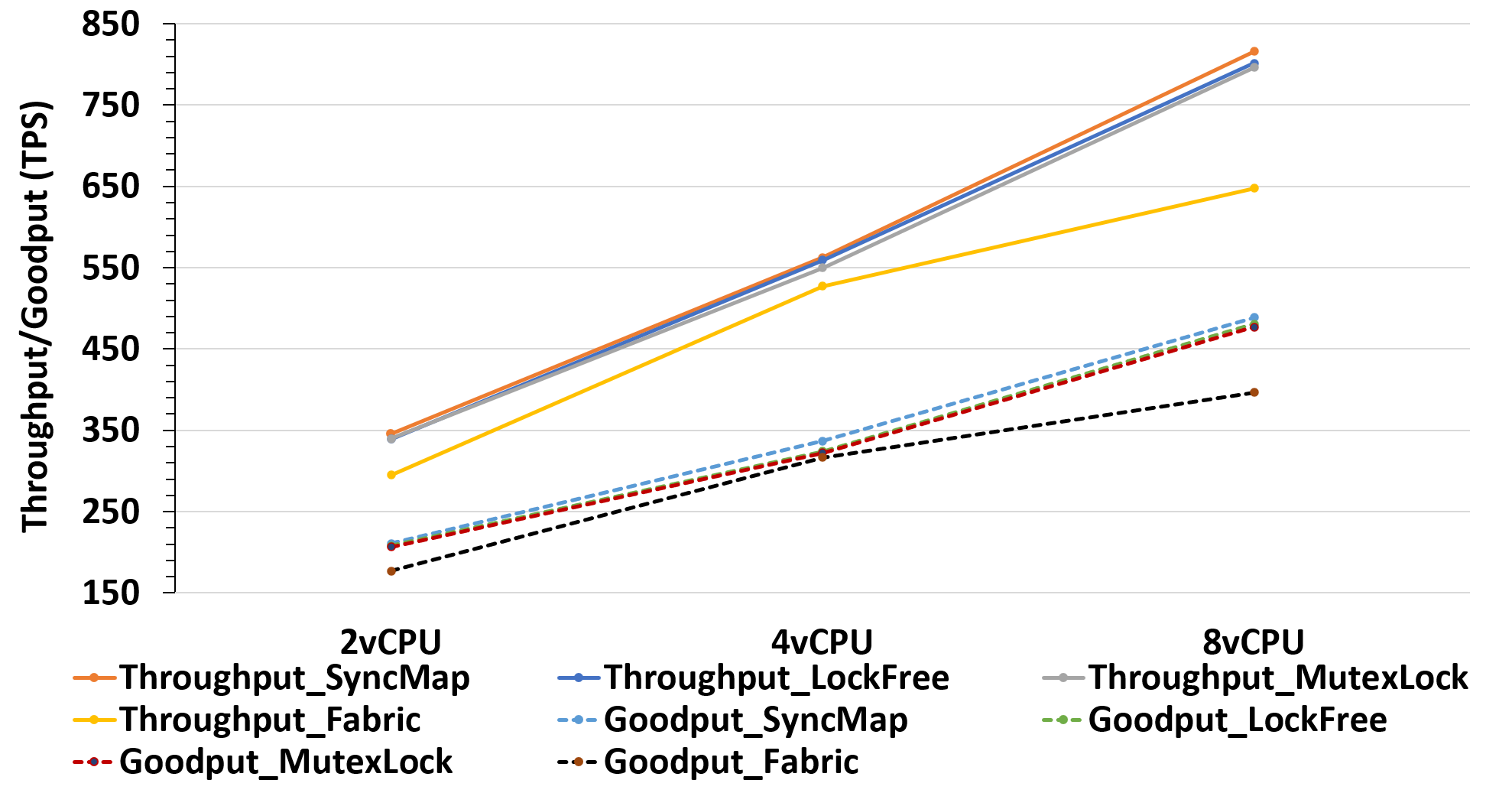}
\caption{vCPUs vs. Throughput} 
\label{fig:figA}
\end{subfigure}
\begin{subfigure}[t]{0.30\textwidth}
\centering
\includegraphics[width=\linewidth]{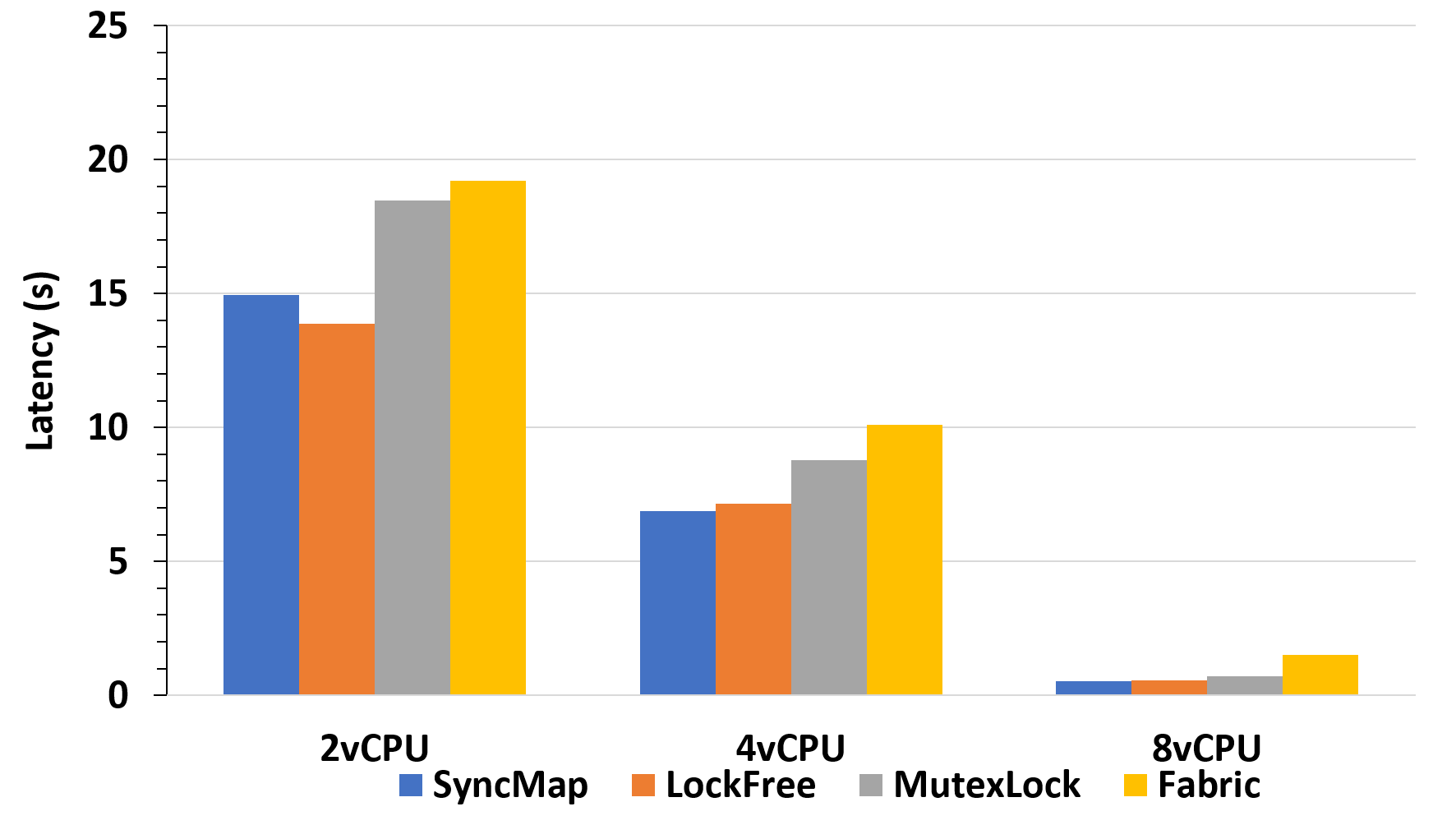}
\caption{vCPUs vs. Latency} 
\label{fig:figB}
\end{subfigure}
\centering
\begin{subfigure}[t]{0.30\textwidth}
\centering
\includegraphics[width=\linewidth]{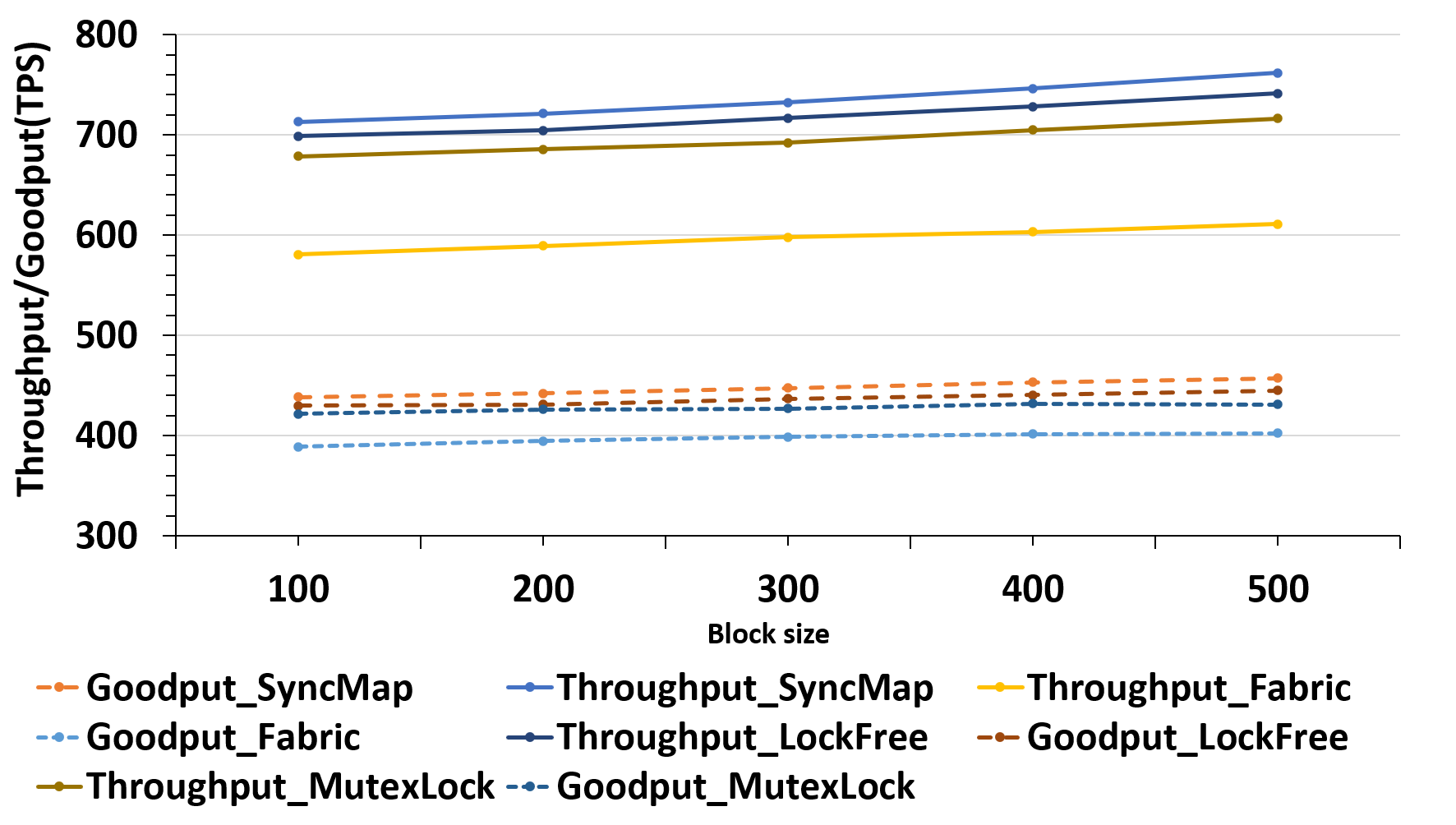}
\caption{Block Size vs. Throughput} 
\label{fig:figC}
\end{subfigure}
\begin{subfigure}[t]{0.30\textwidth}
\centering
\includegraphics[width=\linewidth]{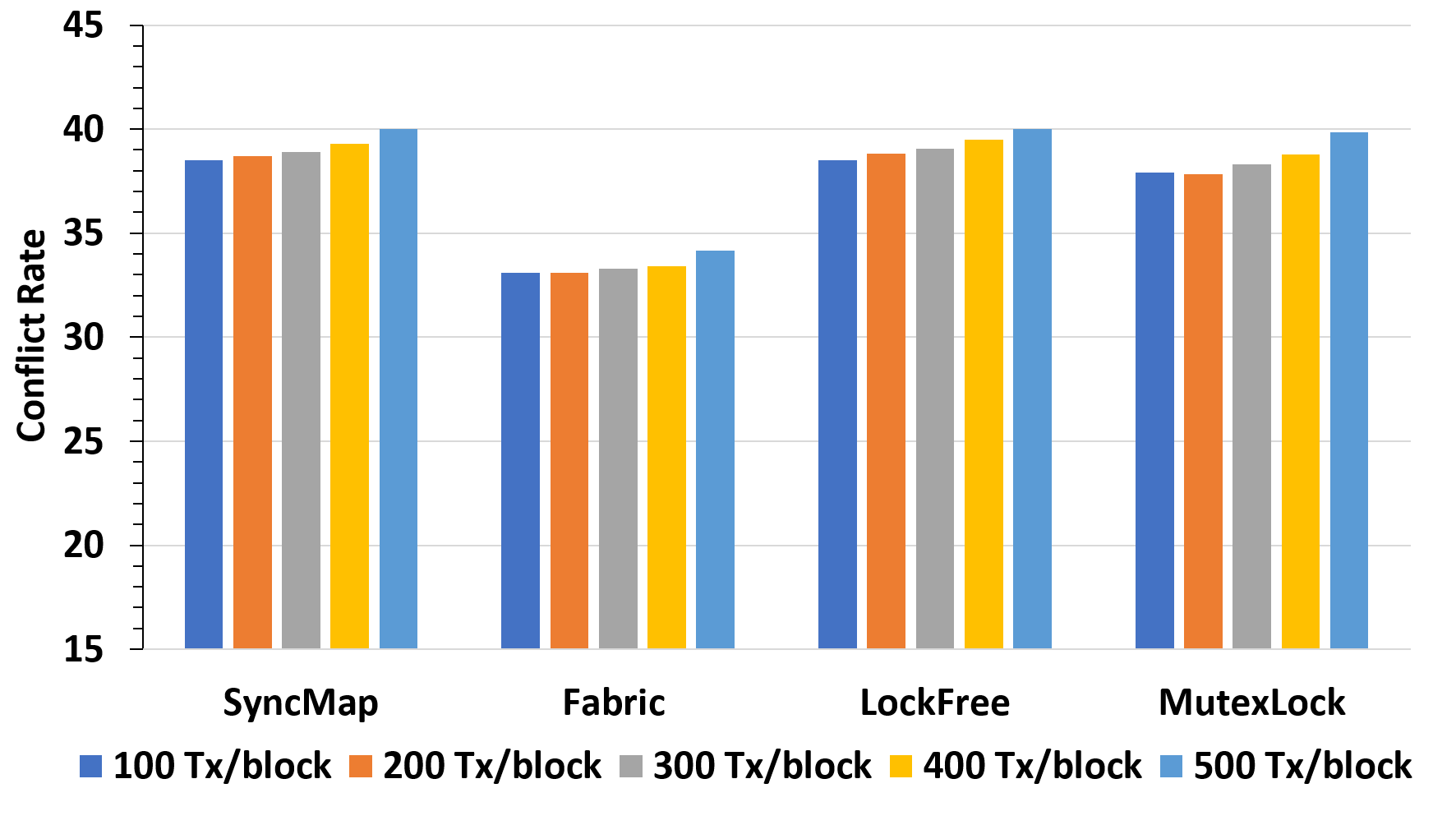}
\caption{Block Size vs. Conflict Rate} 
\label{fig:figD}
\end{subfigure}
\centering
\begin{subfigure}[t]{0.30\textwidth}
\centering
\includegraphics[width=\linewidth]{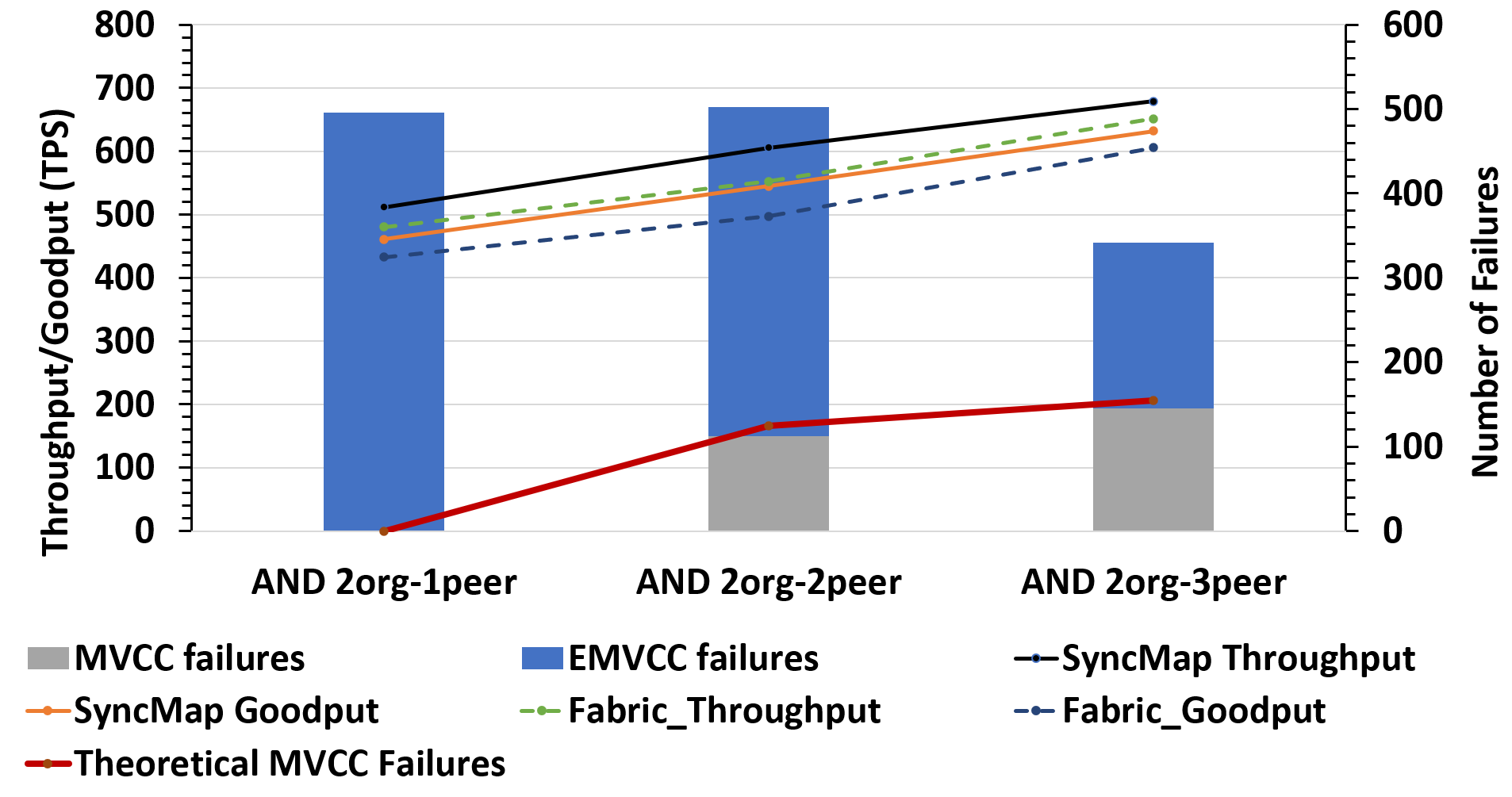}
\caption{Impact of AND Policy} 
\label{fig:figE}
\end{subfigure}
\begin{subfigure}[t]{0.30\textwidth}
\centering
\includegraphics[width=\linewidth]{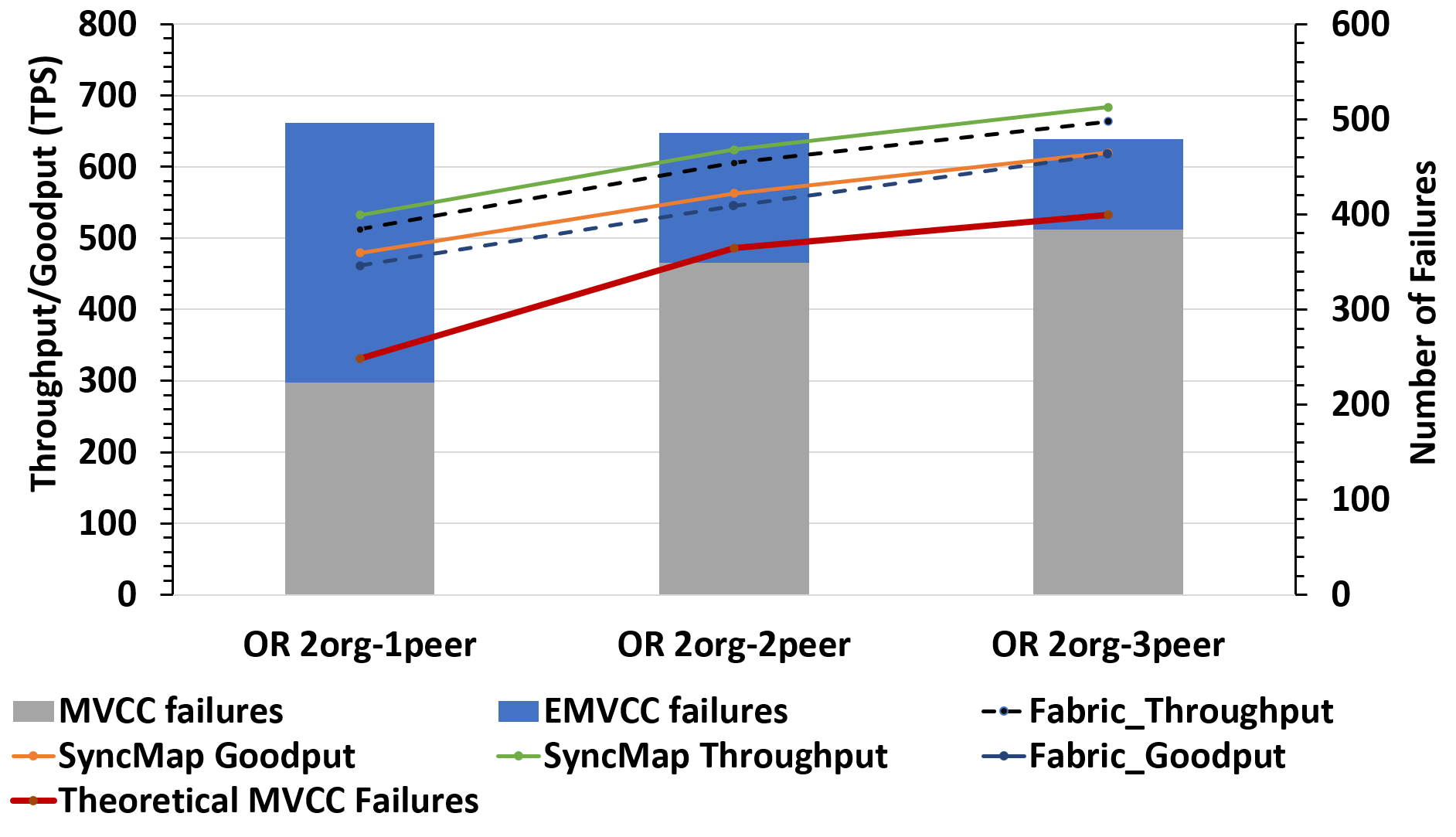}
\caption{Impact of OR Policy} 
\label{fig:figF}
\end{subfigure}
\begin{subfigure}[t]{0.30\textwidth}
\centering
\includegraphics[width=\linewidth]{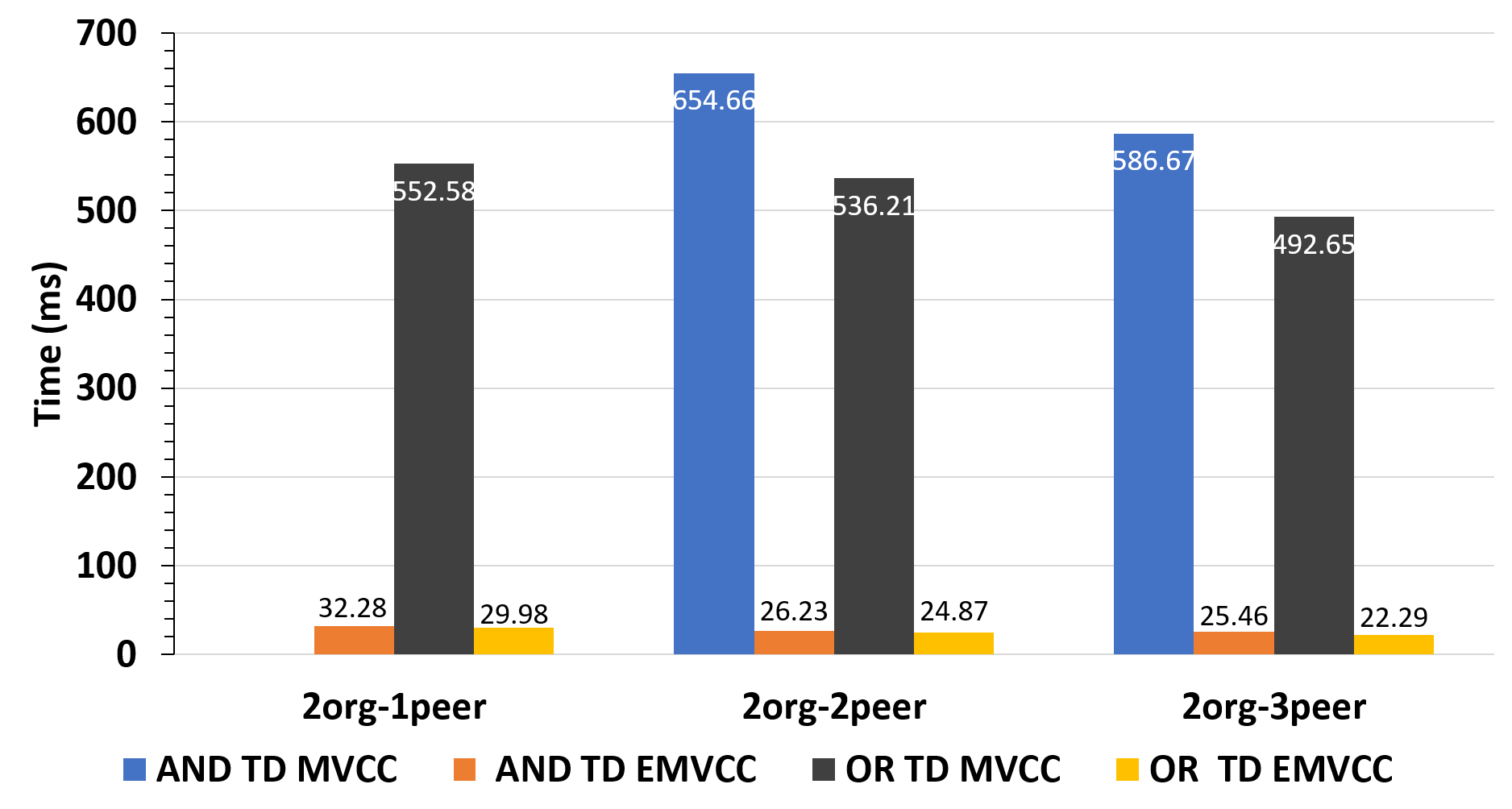}
\caption{Time To Detect EMVCC/MVCC} 
\label{fig:figX}
\end{subfigure}
\centering
\begin{subfigure}[t]{0.30\textwidth}
\centering
\includegraphics[width=\linewidth]{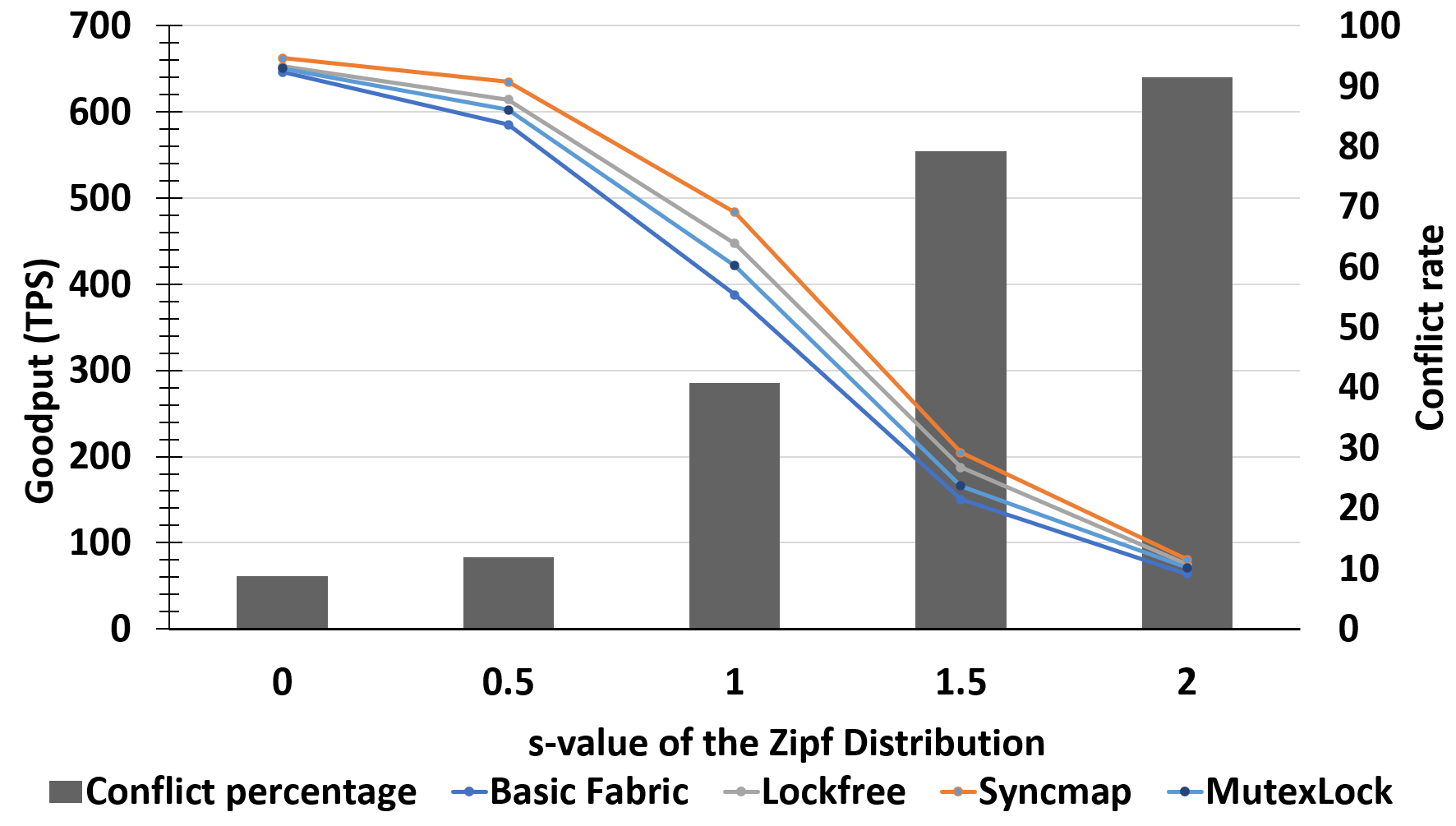}
\caption{Conflict Skewness} 
\label{fig:zipf}
\end{subfigure}
\caption{Sensitivity Analysis Results} 
\end{figure*}

In this part, we will analyze the impact of adding or removing vCPUs to each peer. We vary the number of CPU from 2 vCPUs to 8 vCPUs. Fig.\ref{fig:figA} plots the average throughput and goodput and Fig.\ref{fig:figB} plots the average latency for our three implementations and the Hyperledger Fabric baseline for various vCPU numbers per peer. With an increase in the number of vCPUs per peer, the goodput increases and the latency decreases. Our three solutions are better than the baseline at varying numbers of vCPUs. However, we observe that with 8 vCPUs, there is a significant improvement of 23\% for goodput and 65\% for latency between SyncMap and the baseline. Thus, we conclude that our solutions are able to better leverage additional computational resources than the Fabric baseline.

\subsubsection{Impact of Block Size}
We analyse the impact of the block size by varing the block size from 100 transactions per block to 500 transactions per block.  Fig.\ref{fig:figC} plots the average throughput and goodput for Fabric and our three implementations over various block sizes. Increasing the block size increases the throughput and goodput for all solutions because the use of larger blocks will cause less overhead and fewer network communications. The SyncMap method is offering an average improvement of 12\% compared to Fabric over different block sizes which confirms that our proposed solutions are scalable with different block sizes. Fig.\ref{fig:figD} plots the conflict rate for Fabric and our three methods over various block sizes. We can observe that with an increase in the block size the conflict rate increases: this is due to the increase of the number of intra-block conflicts when the block contains a higher number of transactions. One other observation that can be made is that the conflict rate for Fabric is less than our proposed implementation: this is because there are some false positives and negatives caused by the early MVCC detector.

\subsubsection{Impact of Endorsement Policies and Network Topology}
For this experiment, we use different network topologies by varying the number of peers per organization from one peer to three peers. Fig.\ref{fig:figE} and Fig.\ref{fig:figF} plot the average throughput and goodput, as well as the percentage of transaction failure, caused by EMVCC and MVCC validation for different network topologies using AND (Org1, Org2) and OR (Org1, Org2) endorsement policies for the Fabric baseline and SyncMap solutions. We observe that the type of endorsement policy impacts the network performances. As shown in the two figures, the throughput for the OR endorsement policy is higher than the throughput using the AND policy because using OR the transaction needs less peer endorsement than the AND. For the percentage of non-detection, these experimental results confirm the theoretical results presented in Section \ref{OURSOLUTION}. For example, with AND Policy and using a network with two organizations each one has two peers, we have 22\% of conflicting transactions that were not detected at the EMVCC and were aborted at the MVCC validation (false negative). We can also see that increasing the number of peers increases throughput and goodput for both endorsement policies as the organizations have more resources to process transactions. We can also observe that the increase in the number of peers per organization increases the rate of false positives. The improvement when the AND policy is used is more important compared to Fabric. However, with OR policy, the number of false positives increases which impacts the solution’s performance. We can conclude that with an increase in the number of endorsers our solution becomes more efficient.

Fig.\ref{fig:figX} plots the average time to detect EMVCC/MVCC over different network topologies using OR and AND endorsement policies. The duration of the transaction execution is reduced by 95\% between aborting a conflicting transaction at the EMVCC phase and processing it until reaching the MVCC validation phase to be rejected that is why our solutions reduce significantly the transaction latency. We note that when the number of peers per organization increases, the time to detect MVCC decreases due to the availability of computational resources.

\subsubsection{Impact of Chaincode Implementation}

To simulate another chaincode behavior, we use the Zipf distribution, which allows us to choose the keys used to simulate transactions by varying the parameter \textit{s} of the distribution from 0 to 2. By increasing \textit{s}, we are increasing the preference to use certain same keys which increase the conflict rate. Fig.\ref{fig:zipf} plots the average goodput for the Fabric, SyncMap, LockFree, and MutexLock solutions over different values of the parameter \textit{s} of the Zipf distribution. We can see that when \textit{s} increases the conflict rate increases causing the goodput to decrease. As expected, when the s parameter increases our solutions perform better than Fabric. For example, SyncMap improves the goodput by 24\% for \textit{s} equal to 1 and 35\% for \textit{s} equal 1.5.

\subsection{Summary and Discussions of Findings} 
Our three proposed solutions perform better than the Fabric baseline. SyncMap is the best data structure to implement the cache for keys storing. The improvement rate compared to Fabric is mostly determined by the used endorsement policies, the network design, and the chaincode implementation (conflict rate).  In realistic scenarios, where 40\% of the transactions failed due to MVCC conflict, the SyncMap solution improves the throughput by 23\% and reduces the latency by 80\%.

When using our solution, we recommend developers to use the AND endorsement policy with a maximum number of organizations and the minimum number of peers per organization if the business logic allows that. However, when they have to use the OR endorsement policy, we recommend that they use the minimum number of organizations and number of peers per organization. The block size should be adjusted carefully to minimize the inter-block and intra-block transactions conflict in order to maximize the goodput. For the chaincode implementation, it is important to write a chaincode that ensures a conflict rate lower than 40\%. This solutions can be useful to optimize the performances of other blockchain using the EOV approach. 

\section{CONCLUSIONS}\label{CONCLUSIONS}
In this work, we propose a mechanism to improve Hyperledger Fabric performances aiming to early abort transactions that have no chance to be validated and committed to the ledger using different caching techniques. In an experimental evaluation, we compare SyncMap, LockFree, and MutexLock to the basic version of Fabric. We also did a sensitivity analysis by varying configurable parameters such as block size, endorsement policies, and resource allocation. We show that the proposed solutions outperform Fabric and SyncMap which is the best-proposed method that improves the throughput by up to 20\% and reduces the latency by up to 80\% compared to Fabric. 

\bibliographystyle{unsrt}

\bibliography{references.bib}

\appendix
In this appendix we will provide more details about how we calculate equations:

For equation \ref{eq1}, we can calculate the false positive as follow:
\begin{equation*}
\resizebox{0.9\hsize}{!}{$%
Tx\_invalid\_D = Tx/Bloc \times \%cfct \times (1-\mathcal{P}(ND))
$%
}%
\end{equation*}

\begin{equation*}
\resizebox{0.9\hsize}{!}{$%
Tx\_passing\_endor = Tx\_valid +Tx\_invalid\_ND
$%
}%
\end{equation*}

\begin{equation*}
\resizebox{0.9\hsize}{!}{$%
FP = Tx\_passing\_endor \times \frac{Tx\_invalid\_ND}{Tx\_invalid\_D}
$%
}%
\end{equation*}

\begin{equation*}
\resizebox{0.9\hsize}{!}{
	$FP= \frac{{Tx/Bloc} \times \%cfct^2 \times \mathcal{P}(ND) \times (1-\mathcal{P}(ND))} {\%cfct \times \mathcal{P}(ND) - \%cfct + 1}$
}%
\end{equation*}
where $Tx\_invalid\_D$ is the number of invalid transactions detected, $Tx\_invalid\_ND$ is the number of valid transactions non-detected and $Tx\_passing\_endor$ is the number of transactions passing the endorsement phase

For equation \ref{eq2} we calculate the false positive as follow:

\begin{align}
    FN = Nb\_cfct\_TX\ \times \mathcal{P}(ND)\nonumber\\
     FN = {Tx/Bloc} \times \%cfct \times \mathcal{P}(ND) \nonumber
 \end{align}
 
where the $Nb\_cfct\_TX$ is the number of transactions in conflict. 

For equation \ref{eq4}, the probability of non-detection using the AND endorsement policy is defined as follow: let's suppose that we have two transactions (\textit{TX1} and \textit{TX2}), the client submits \textit{TX1} which is endorsed by a peer of each organization on the network. Then a second transaction \textit{TX2} having an MVCC conflict with \textit{TX1} was submitted. The probability of non-detection is the probability that the transaction \textit{TX2} is not endorsed by a peer who has endorsed the transaction \textit{TX1}. The probability of non-detection is the product of the conditional probabilities of non-detection of each organization. The probability of non-detection knowing that a peer belongs to an organization is the number of peers who have not endorsed \textit{TX1} which is ($ M_i-1 $) divided by the total number of peers of this organization ($ M_i $).

\begin{align}
\mathcal{P}(ND) = \prod_{i=1}^{N} {P}(ND \mid Org_i) \nonumber\\
    \mathcal{P}(ND) = \prod_{n=1}^{N} \frac{(M_i-1)}{M_i}\nonumber
\end{align}

Finally, for \ref{eq5} the probability of non-detection using the OR endorsement policy is calculated as follow: if all organizations have the same number of peers, the probability of no detection is the result of substituting one minus the probability of detection, and for M peer per organization, the probability of detection is $ \ frac {1} {NM} $.
\begin{align}
\mathcal{P}(ND) = 1-\mathcal{P}(D) \nonumber\\
\mathcal{P}(ND) = 1- \frac{1}{NM} \nonumber\\
    \mathcal{P}(ND) = \frac{NM-1}{NM}\nonumber
\end{align}

Otherwise, if the organizations have different numbers of peers, the probability of non-detection is calculated as follows: it is the sum of the products of the probability that the transaction will be endorsed by an organization and the conditional probability of non-detection by organization. The probability of having an endorsement from a particular peer is $\frac{1}{N}$ and the conditional probability of not being detected by the organization is the probability of receiving an endorsement from a peer that has not endorsed the first transaction which is equivalent to $ \frac{1}{M_i} $.
\begin{align}
\mathcal{P}(ND) = \sum_{i=1}^N \mathcal{P}(Org_i) \times {P}(ND \mid Org_i) \\
\mathcal{P}(ND) = \frac{1}{N} \sum_{i=1}^N \frac{1}{C^{M_i-1}_{M_i}} \\
\mathcal{P}(ND) = \frac{1}{N} \sum_{i=1}^N \frac{1}{M_i}
\end{align}

If we summarize, we obtain the formula for calculating the probability of non-detection using the OR endorsement policy defined in equation \ref{eq5}.

\end{document}